\documentstyle[12pt]{article}
\input epsf
\newlength{\dinwidth}
\newlength{\dinmargin}
\begin{document}

\def\ap#1#2#3   {{\em Ann. Phys. (NY)} {\bf#1} (#2) #3.}   
\def\apj#1#2#3  {{\em Astrophys. J.} {\bf#1} (#2) #3.} 
\def\apjl#1#2#3 {{\em Astrophys. J. Lett.} {\bf#1} (#2) #3.}
\def\app#1#2#3  {{\em Acta. Phys. Pol.} {\bf#1} (#2) #3.}
\def\ar#1#2#3   {{\em Ann. Rev. Nucl. Part. Sci.} {\bf#1} (#2) #3.}
\def\cpc#1#2#3  {{\em Computer Phys. Comm.} {\bf#1} (#2) #3.}
\def\err#1#2#3  {{\it Erratum} {\bf#1} (#2) #3.}
\def\ib#1#2#3   {{\it ibid.} {\bf#1} (#2) #3.}
\def\jmp#1#2#3  {{\em J. Math. Phys.} {\bf#1} (#2) #3.}
\def\ijmp#1#2#3 {{\em Int. J. Mod. Phys.} {\bf#1} (#2) #3}
\def\jetp#1#2#3 {{\em JETP Lett.} {\bf#1} (#2) #3.}
\def\jpg#1#2#3  {{\em J. Phys. G.} {\bf#1} (#2) #3.}
\def\mpl#1#2#3  {{\em Mod. Phys. Lett.} {\bf#1} (#2) #3.}
\def\nat#1#2#3  {{\em Nature (London)} {\bf#1} (#2) #3.}
\def\nc#1#2#3   {{\em Nuovo Cim.} {\bf#1} (#2) #3.}
\def\nim#1#2#3  {{\em Nucl. Instr. Meth.} {\bf#1} (#2) #3.}
\def\np#1#2#3   {{\em Nucl. Phys.} {\bf#1} (#2) #3}
\def\pcps#1#2#3 {{\em Proc. Cam. Phil. Soc.} {\bf#1} (#2) #3.}
\def\pl#1#2#3   {{\em Phys. Lett.} {\bf#1} (#2) #3}
\def\prep#1#2#3 {{\em Phys. Rep.} {\bf#1} (#2) #3.}
\def\prev#1#2#3 {{\em Phys. Rev.} {\bf#1} (#2) #3}
\def\prl#1#2#3  {{\em Phys. Rev. Lett.} {\bf#1} (#2) #3}
\def\prs#1#2#3  {{\em Proc. Roy. Soc.} {\bf#1} (#2) #3.}
\def\ptp#1#2#3  {{\em Prog. Th. Phys.} {\bf#1} (#2) #3.}
\def\ps#1#2#3   {{\em Physica Scripta} {\bf#1} (#2) #3.}
\def\rmp#1#2#3  {{\em Rev. Mod. Phys.} {\bf#1} (#2) #3}
\def\rpp#1#2#3  {{\em Rep. Prog. Phys.} {\bf#1} (#2) #3.}
\def\sjnp#1#2#3 {{\em Sov. J. Nucl. Phys.} {\bf#1} (#2) #3}
\def\spj#1#2#3  {{\em Sov. Phys. JEPT} {\bf#1} (#2) #3}
\def\spu#1#2#3  {{\em Sov. Phys.-Usp.} {\bf#1} (#2) #3.}
\def\zp#1#2#3   {{\em Zeit. Phys.} {\bf#1} (#2) #3}

\title{\vspace{3cm}
\bf{ The transition from the photoproduction to the DIS region }\thanks{Talk 
presented at the DIS 96 Workshop, Rome, 15-19 April, 1996.}
\vspace{2cm}}

\author{
 {\bf Aharon Levy} \\ 
{\small \sl School of Physics and Astronomy}\\ {\small \sl Raymond and 
Beverly Sackler Faculty of Exact Sciences}\\
  {\small \sl Tel--Aviv University, Tel--Aviv, Israel}
}
\date{ }
\maketitle

\vspace{5cm}

\begin{abstract}
This is a review of the existing ideas in modelling the transition 
from the photoproduction ($Q^2 = 0$) to the deep inelastic 
scattering region. The available data of the proton structure 
function are analyzed from the point of view of the total $\gamma^* p$ 
cross section behavior with $W$ and the results are compared to 
different parameterizations. The question whether DIS processes are hard 
or soft is discussed.
\end{abstract}

\vspace{-20cm}
\begin{flushright}
TAUP 2349-96 \\
July 1996 \\
\end{flushright}

\setcounter{page}{0}
\thispagestyle{empty}
\newpage  

\section{Introduction}

One of the aims of building HERA was to study the deep inelastic
scattering (DIS) region with data at low $x$ and high $Q^2$.  Yet,
recently efforts are being made to get to lower and lower $Q^2$ values in
the low $x$ region in order to study the transition from photoproduction
to the DIS regime. The main motivation for looking at the transition
region is the following: at $Q^2=0$ the dominant processes are of
non--perturbative nature and are well described by the Regge picture. As
$Q^2$ increases, the exchanged photon is expected to shrink and one
expects perturbative QCD to take over. What can one learn from the
transition between soft processes (low virtuality) and hard processes
(high virtuality)? Where does the change take place? Is it a sudden 
transition or a smooth one? The transition should shed light on the 
interplay between soft and hard interactions.
In addition, being able to describe this region has
a practical necessity:  it is needed for calculating radiative
corrections. 

The purpose of this talk is to review the ideas in modelling 
the low $x$ and low $Q^2$ region~\cite{bkrev}. 

\section{The models}

\subsection{Donnachie and Landshoff (DL) }

Donnachie and Landshoff~\cite{dl1} found a simple Regge picture 
describing all hadron--hadron cross sections with a sum of two terms, 
that of a pomeron exchange and that of a reggeon. They showed this 
picture to describe also real photoproduction cross sections. They 
extended the picture for virtual photons ($\gamma^*, Q^2 <$ 10 GeV$^2$) to 
see what is the expected contribution of the non--perturbative mechanism 
to higher $Q^2$~\cite{dl2}. The main interest is in the low $x$ region where 
the pomeron dominates and thus the question of interest is what is the 
contribution of the `soft' pomeron at intermediate $Q^2$.

\subsection{Capella, Kaidalov, Merino, Tran--Than--Van (CKMT) }

In this picture~\cite{ckmt} there is no `soft' or `hard' pomeron, there is 
just one `bare' pomeron. At low $Q^2$ absorptive corrections 
(rescattering) give a pomeron with an effective intercept of $ 1 + 
\Delta_0 (\Delta_0 \sim 0.08)$. If one uses an eikonal approach, the bare 
intercept becomes $ 1 + \Delta_1 (\Delta_1 \sim 0.13)$. A more complete 
absorptive calculation results in $ 1 + \Delta_2 (\Delta_2 \sim 0.24)$. 
The absorptive corrections decrease rapidly with $Q^2$. They 
parametrize the data with this behavior of the pomeron
up to $Q^2 <$ 5 GeV$^2$ and use it then as initial conditions to a pQCD 
evolution.

\subsection{ Badelek and Kwiecinski (BK) }

Badelek and Kwiecinski~\cite{bk} describe the low $Q^2$ region by using 
the generalized vector dominance model (GVDM): the proton 
structure function $F_2$ is represented by the contribution of a large 
number of vector mesons which couple to virtual photons. The low mass 
ones, $\rho, \omega, \phi$ contribute mainly at low $Q^2$, while the 
higher mass are determined by the asymptotic structure function 
$F_2^{AS}$ which is described by pQCD. The total structure function is 
given by a $Q^2$ weighted sum of the two components.

\subsection{Abramowicz, Levin, Levy, Maor (ALLM) }

This parameterization~\cite{allm} is based on a Regge motivated approach
extended into the large $Q^2$ regime in a way compatible with QCD
expectations. This approach allows to parametrize the whole $x, Q^2$ phase
space, fitting all the existing data. 

\subsection{Some general comments}

The DL parametrization provides a good way to check to what value of $Q^2$ 
can the simple 'soft' pomeron picture be extended. It is not meant to be 
a parameterization which describes the whole DIS regime. The CKMT and BK 
parametrizations are attempts to get the best possible presentation of the 
initial conditions to a pQCD evolution. The ALLM does not use the regular 
pQCD evolution equation but parametrizes the whole of the DIS phase 
space by a combination of Regge and QCD motivated parametrizations.

All parameterizations make sure that as $Q^2 \to 0$ also $F_2 \to 0$ 
linearly with $Q^2$.

\section{Details of the parametrizations}

\subsection{The DL parameterization}

The proton structure function $F_2$ is given by 
\begin{equation}
F_2(x,Q^2) \sim A \xi^{-0.0808} \phi(Q^2) + B \xi^{0.4525} \psi(Q^2) ,
\end{equation}
where $\xi$ is the rescaled variable
\begin{equation}
\xi = x \left( 1 + \frac{\mu^2}{Q^2} \right) ,
\end{equation}
with $x$ being the Bjorken--$x$ and the scale variable $\mu$ has 
different values for different flavors: for $u$ and $d$ quarks $\mu = 
0.53$ GeV, for the strange quark $s$, $\mu = 1.3$ GeV and for the charm 
quark $c$, $\mu = 2$ GeV. The two functions $\phi(Q^2)$ and $\psi(Q^2)$ 
make sure that the structure function vanishes linearly with $Q^2$ as 
$Q^2 \to$ 0,
\begin{equation}
\phi(Q^2) = \frac{Q^2}{Q^2 + a} \ \ \ \ \ \ \psi(Q^2) = \frac{Q^2}{Q^2 + b} .
\end{equation}
The four parameters $A, B, a$ and $b$ are constrained so as to reproduce 
the total real photoproduction data,
\begin{equation}
\frac{A}{a} (\mu^2)^{-0.0808} = 0.604 \ \ \ \ \ \frac{B}{b} 
(\mu^2)^{0.4525} = 1.15 .
\end{equation}
In addition there is also a higher--twist term $ht(x,Q^2)$ contributing to 
the structure function,
\begin{equation}
ht(x,Q^2) = D \frac{x^2(1 - \xi)^2}{1 + \frac{Q^2}{Q_0^2}} ,
\end{equation}
with the parameters $D$ = 15.88 and $Q_0$ = 550 MeV.

\subsection{The CKMT parameterization}

Contrary to the DL parameterization, the CKMT assumes that the power 
behavior of $x$ is $Q^2$ dependent,
\begin{equation}
F_2(x,Q^2) = A x^{-\Delta(Q^2)} (1 - x)^{n(Q^2)+4} \left( \frac{Q^2}{Q^2 
+ a} \right)^{1+\Delta(Q^2)} + B x^{1-\alpha_R} (1 - x)^{n(Q^2)} \left( 
\frac{Q^2}{Q^2 + b} \right)^{\alpha_R} ,
\end{equation}
where $\alpha_R$ is the Reggeon trajectory intercept,
the power $n(Q^2)$ is given by
\begin{equation}
n(Q^2) = \frac{3}{2} \left( 1 + \frac{Q^2}{Q^2 + c} \right) 
\end{equation}
and the power behavior of $x$ is given by
\begin{equation}
\Delta(Q^2) = \Delta_0 \left( 1 + \frac{Q^2}{Q^2 + d} \right) .
\end{equation}
The constant parameters are determined by the requirement that $F_2$ and 
the derivative $\frac{dF_2}{d\ln Q^2}$ at $Q^2 = Q_0^2$ to coincide with 
that obtained from the pQCD evolution equations. They can do so at 
$Q^2_0$ = 2 GeV$^2$, provided a higher--twist term is added to that of pQCD,
\begin{equation}
F_2(x,Q^2) = F_2^{pQCD}(x,Q^2) \left( 1 + \frac{f(x)}{Q^2} \right)
\end{equation} 
for $ Q^2 \geq Q_0^2$. The values of the parameters are: $A$ = 0.1502, $a$ 
= 0.2631 GeV$^2$, $\Delta_0$ = 0.07684, $d$ = 1.117 GeV$^2$, $b$ = 0.6452 
GeV$^2$, $\alpha_R$ = 0.415, $c$ = 3.5489 GeV$^2$.

\subsection{The BK parameterization}

The proton structure function is written as the sum of two terms, a 
vector meson part (V) and a partonic part (par),
\begin{equation}
F_2(x,Q^2) = F_2^V(x,Q^2) + F_2^{par}(x,Q^2) .
\end{equation}
The part representing the contribution from vector mesons which couple to 
the virtual photon is given by
\begin{equation}
F_2^V(x,Q^2) = \frac{Q^2}{4\pi} \Sigma_V \frac{M_V^4 
\sigma_V(W^2)}{\gamma_V^2(Q^2 + M_V^2)^2} ,
\end{equation}
where $\gamma_V^2/(4\pi)$ is the direct photon vector meson coupling,
$W$ is the $\gamma^* p$ center of mass energy and $\sigma_V$ is the total 
$Vp$ cross section. The sum is over all vector meson satisfying $M_V^2 < 
Q_0^2$, where $M_V$ is the mass of the vector meson and $Q_0$ is a 
parameter. 

The partonic part of the structure function is given by the expression
\begin{equation}
F_2^{par}(x,Q^2) = \frac{Q^2}{Q^2 + Q^2_0} F_2^{AS}(\bar{x},Q^2+Q^2_0) ,
\end{equation}
where the asymptotic structure function $F_2^{AS}$ is given by pQCD at 
the scaled value of
\begin{equation}
\bar{x} = \frac{Q^2 + Q^2_0}{W^2 + Q^2 -M^2 + Q^2_0} ,
\end{equation}
where $M$ is the proton mass. In practice the parameterization uses 
$Q^2_0$ = 1.2 GeV$^2$ and thus sums over the contribution of the 3 
lightest vector mesons $\rho, \omega$ and $\phi$.

\subsection{The ALLM parameterization}

This parameterization attempts to cover the whole of $x,Q^2$ region above 
the resonances ($W^2 >$ 3 GeV$^2$), at the expense of introducing more 
parameters than the other parameterizations. The proton structure 
function has the form
\begin{equation}
F_2(x,Q^2) = \frac{Q^2}{Q^2 + M_0^2} \left( F_2^{\cal P}(x,Q^2) + F_2^{\cal 
R}(x,Q^2) \right) , 
\end{equation}
where $M_0$ is the effective photon mass. The functions $F_2^{\cal P}$ 
and $F_2^{\cal R}$ are the contribution of the 
pomeron $\cal P$ or Reggeon $\cal R$ exchanges to the structure function. 
They take the form
\begin{eqnarray}
F_2^{\cal P}(x,Q^2) &=& c_{\cal P}(t) x_{\cal P}^{a_{\cal P}(t)} (1 - 
x)^{b_{\cal P}(t)} ,  \\ 
F_2^{\cal R}(x,Q^2) &=& c_{\cal R}(t) x_{\cal R}^{a_{\cal R}(t)} (1 - 
x)^{b_{\cal R}(t)} . \nonumber 
\end{eqnarray}
The slowly varying function $t$ is defined as
\begin{equation}
t = \ln \left( \frac{ \ln \frac{Q^2 + Q^2_0}{\Lambda^2}}{\ln 
\frac{Q^2_0}{\Lambda^2}} \right) .
\end{equation}
The two scaled variables $x_{\cal P}$ and $x_{\cal R}$ are modified 
Bjorken--$x$ variables which include mass parameters $M_{\cal P}$ and 
$M_{\cal R}$ which can be interpreted as effective pomeron and reggeon 
masses: 
\begin{eqnarray}
\frac{1}{x_{\cal P}} &=& 1 + \frac{W^2 - M^2}{Q^2 + M_{\cal P}^2} , \\ 
\frac{1}{x_{\cal R}} &=& 1 + \frac{W^2 - M^2}{Q^2 + M_{\cal R}^2} . \nonumber
\end{eqnarray} 

\section{Comparison to data}

The proton structure function has been measured~\cite{eisele} in a wide
range of $x$ and $Q^2$. A constructive way to display them~\cite{wdep}
over a wide kinematical range is to use the relation between $F_2$ and the
total $\gamma^* p$ cross section,
\begin{equation}
\sigma^{\gamma^* p}(W^2, Q^2) = \frac{4\pi^2\alpha}{Q^2} \frac{1}{1 - x} 
\left( 1 + \frac{4 M^2 x^2}{Q^2} \right) F_2(x,Q^2) .
\label{sigf2}
\end{equation}
In this expression the Hand definition of the flux of virtual photons is
used. 
\vspace{0.5cm}

\setlength{\unitlength}{0.7mm}
\begin{figure}[hbtp]
\begin{picture}(100,145)(-70,1)
\mbox{\epsfxsize7.8cm\epsffile{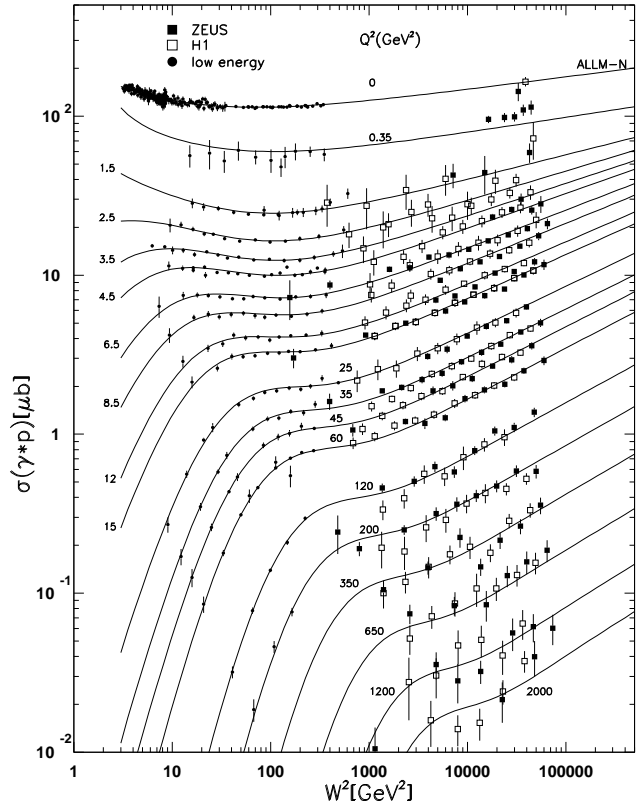}}
\end{picture}
\caption{The total $\gamma^*p$ cross section as function of $W^2$ from the
$F_2$ measurements for different $Q^2$ values. The
lines are the expectations of the ALLM parameterization.}
\label{fig:gstarp_allm}
\end{figure}

Figure \ref{fig:gstarp_allm} presents the dependence of
$\sigma_{tot}^{\gamma^*p}$, obtained through equation \ref{sigf2} from the
measured $F_2$ values~\cite{ZEUSf2,H1f2}, on the square of the center of
mass energy, $W^2$, for fixed values of the photon virtuality $Q^2$.
The new preliminary very low $Q^2$ measurements of the ZEUS 
collaboration~\cite{bpc}, as well as those of the NMC collaboration 
presented at this workshop~\cite{ewa} are included in the figure. 
Also shown are the measurements of the total real photoproduction cross
sections. While the data below $Q^2$=1 GeV$^2$ shows a very mild $W$
dependence, which resembles that of the real photon data, the trend
changes as $Q^2$ increases. Note that for higher values of $Q^2$ one sees
the typical threshold behavior for the case when $W^2<Q^2$~\cite{LM}. The
curves are the results of a new ALLM type parameterization which includes
now also the data from E665~\cite{E665}, those from
NMC~\cite{newNMC} and the published HERA~\cite{pubHERA} data. 

\vspace{0.5cm}
\setlength{\unitlength}{0.7mm}
\begin{figure}[hbtp]
\begin{picture}(100,125)(-70,1)
\mbox{\epsfxsize6.5cm\epsffile{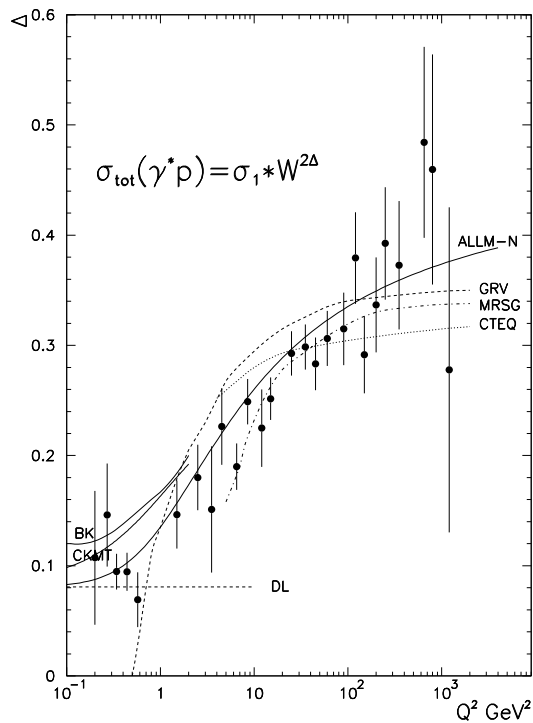}}
\end{picture}
\caption{The $Q^2$ dependence of the parameter $\Delta$ obtained from a fit
of the expression $\sigma_{tot}^{\gamma^*p} = \sigma_1 W^{2\Delta}$ to the
data in each $Q^2$ bin. The curves are the expectations of the
parameterizations mentioned in the text.}
\label{fig:delta}
\end{figure}

Instead of comparing the data as presented in figure \ref{fig:gstarp_allm}
with the different parameterizations, it is more economical as well as
instructive to study the energy dependence of the $\gamma^* p$ cross
section for fixed $Q^2$ values~\cite{AL}.  In order to see how the slope
of the $W$ dependence changes with $Q^2$, the cross section values in the
region where $W^2\gg Q^2$ were fitted to the form
$\sigma_{tot}^{\gamma^*p} = \sigma_1 W^{2\Delta}$ for each fixed $Q^2$
interval. The resulting values of $\Delta$ from the fit are plotted
against $Q^2$ in figure \ref{fig:delta}. Similar results have been obtain
by the H1 collaboration~\cite{H1f2} who use only their own data to fit
the structure function measurements to the form $F_2 \sim x^{-\Delta}$.
Also included in the figure are the recent preliminary results of the 
ZEUS collaboration~\cite{bpc} in the region $ 0.2 < Q^2 < 0.8$ GeV$^2$.
One can see the slow increase of $\Delta$ with $Q^2$ from the value of
0.08 at $Q^2$=0, to around 0.2 for $Q^2 \sim$ 10--20 GeV$^2$ followed by a
further increase to around 0.3--0.4 at high $Q^2$. One would clearly
profit from more precise data, expected to come soon. 
 
The curves are the expectations of the DL, BK, CKMT, and the updated ALLM
parameterization, which includes also some of the recent HERA data in its
fit. In addition, the expectations of the GRV~\cite{grv}, CTEQ~\cite{cteq}
and MRSG~\cite{mrsg} are also shown. 

The DL parameterization can describe the data up to $Q^2 \sim$ 1 GeV$^2$. 
All the others give in general the right features of the $Q^2$ behavior
with a smooth transition from soft to hard interactions with an interplay
between the two in the intermediate $Q^2$ range. 

\section{DIS processes - hard or soft?}

What have we learned from the behavior of the data with $Q^2$? What are 
we actually measuring? At low $Q^2$ the photon is known to have 
structure. Does $F_2$ still measure the structure of the proton? 
Bjorken~\cite{bj94} pointed out that physics is not frame dependent. The 
structure of the proton alone has no meaning. One has to study the 
$\gamma^* p$ interaction.

Let us look at the structure of a photon. It is a well known fact that
real photon behave like hadrons when interacting with other hadrons.
One way to understand this is by using the argument of
Ioffe~\cite{ioffe}: the photon can fluctuate into a $q\bar{q}$ pair.
The fluctuation time is given by
\begin{equation}
t_f = \frac{2 E_{\gamma}}{m_{q\bar{q}}^2}
\end{equation}
where $E_{\gamma}$ is the photon energy in the rest system of the proton and 
$m_{q\bar{q}}$ is the mass of the $q\bar{q}$ system into which the photon 
fluctuates. 
The Vector Dominance Model assumes that the fluctuation of the photon is into 
vector mesons, $m_{q\bar{q}} \simeq m_V$, where $m_V$ is the vector mesom 
mass.
As long as $t_f\gg t_i$, where the interaction time $t_i \approx r_p$, with 
$r_p$ being the proton radius, the photon interacts as if it were a hadron. 
When the photon becomes virtual with a negative square mass of 
$Q^2$, its fluctuation time becomes
\begin{equation}    
t_f = \frac{2 E_{\gamma}}{{m_{q\bar{q}}^2} + Q^2}
\end{equation}
and thus at low energies and moderate Bjorken $x$, the fluctuation time 
becomes small and the virtual photon behaves like a point--like structureless 
object, consistent with the DIS picture described above.

However, at high energies or equivalently in the low $x$ region studied at 
HERA, the fluctuation time of a virtual photon can be expressed as
\begin{equation}
t_f \approx \frac{1}{2 M x}
\end{equation}
where $M$ is the proton mass. This can be derived easily from formula 
(20), assuming ${m_{q\bar{q}}^2} \approx Q^2$~\cite{afs}.
Thus in the HERA regime, a photon of 
virtuality as high as $Q^2 \sim 2-3 \times 10^3$ GeV$^2$ can fluctuate into a 
$q\bar{q}$ pair, which will survive till arrival on the proton target.

The photon can fluctuate into typically two configurations. A large size
configuration will consist of an asymmetric $q\bar{q}$ pair with each
quark carrying a small transverse momentum $k_T$ (fig. \ref{fig:ajm}(a)).
For a small size configuration the pair is symmetric, each quark having a
large $k_T$ (fig. \ref{fig:ajm}(b)). One expects the asymmetric large
configuration to produce 'soft' physics, while the symmetric one would
yield the 'hard' interactions. 

In the aligned jet model (AJM)~\cite{ajm} the first 
configuration dominates while the second one is the 'sterile combination' 
because of color screening. In the photoproduction case ($Q^2$ = 0), the 
small $k_T$ configuration dominates. Thus one has large color forces 
which produce the hadronic component, the vector mesons, which finally 
lead to hadronic non--perturbative final states of 'soft' nature. The 
symmetric configuration contributes very little. In those cases 
where the photon does fluctuate into a high $k_T$ pair, color 
transparency suppresses their contribution.

\vspace{0.5cm} 
\setlength{\unitlength}{0.7mm}
\begin{figure}[hbtp]
\begin{picture}(150,40)(-50,90)
\mbox{\epsfxsize10.0cm\epsffile{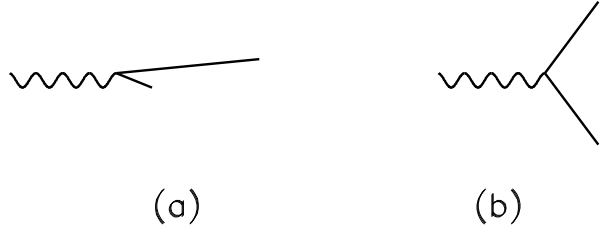}}
\end{picture}
\caption{Fluctuation of the photon into a $q \bar{q}$ pair in (a)
asymmetric small $k_T$ configuration, (b) into a symmetric large $k_T$
configuration}
\label{fig:ajm}
\end{figure}   

In the DIS regime ($Q^2 \neq$ 0), the symmetric contribution becomes bigger. 
Each such pair still contributes very little because of color 
transparency, but the phase space for the symmetric configuration 
increases. However the asymmetric pair still contribute also to the DIS 
processes. In fact, in the quark parton model (QPM) the fast quark 
becomes the current jet and the slow quark interacts with the proton 
remnant resulting in processes which look in the $\gamma^* p$ frame just 
like the 'soft' processes discussed in the $Q^2$ = 0 case. So there 
clearly is an interplay between soft and hard interactions also in the 
DIS region.

This now brings up another question. We are used by now to talk about the 
'resolved' and the 'direct' photon interactions. However,
if the photon always fluctuates into a $q \bar{q}$ pair even at quite 
large values of $Q^2$, what does one mean by a 'direct' photon 
interaction? To illustrate the problem, let us look at the diagram 
describing the photon--gluon fusion, which is usually considered in 
leading order a direct photon interaction and is shown in figure 
\ref{fig:direct}(a). An example of a resolved process is shown in figure 
\ref{fig:direct}(b) where a photon fluctuates into a $q \bar{q}$ pair 
with a given $k_T$, following by the interaction of one of the quarks 
with a gluon from the proton to produce a quark and a gluon with a given 
$p_T$.

\vspace{0.5cm}
\setlength{\unitlength}{0.7mm}
\begin{figure}[hbtp]
\begin{picture}(150,60)(-50,90)
\mbox{\epsfxsize10.0cm\epsffile{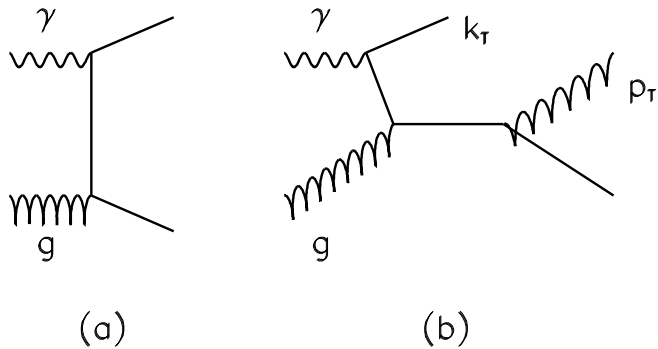}}
\end{picture}
\caption{Diagrams describing examples of (a) 'direct' photon process, (b) 
'resolved' photon process}
\label{fig:direct}
\end{figure}

In the diagram shown in figure \ref{fig:direct}(b) there are two scales, 
$k_T$ and $p_T$. The classification of the process as 'direct' or 
'resolved' depends on the relations between the two scales. If $k_T \ll 
p_T$ we call it a resolved photon interaction, while in the case of $k_T 
\gg p_T$ one would consider this as a direct photon interaction. 
Practically in the latter case the $p_T$ is too small to resolve the 
gluon and the quark jets as two separate jets, thus making it look like 
the diagram in figure \ref{fig:direct}(a). At low $Q^2$ the more likely 
case is that of $k_T \ll p_T$ and thus the resolved photon is the 
dominant component, while at high $Q^2$ the other case is more likely. A 
yet open question is how does one deal with the case where $k_T \sim p_T$.

\section{Summary}

The energy behavior of the $\gamma^* p$ cross section shows that there 
is a smooth transition between the $Q^2$ region where there is a mild energy 
dependence to that where the energy behavior is steeper. It happens 
somewhere in the region of about 1 GeV$^2$. Does this tell us where soft 
interactions turn into hard ones? In order to understand the structure of 
the dynamics, one has to isolate in the transition region the specific 
configurations in $k_T$ and $p_T$ for a better insight of what is 
happening. 

\section*{Acknowledgments}
 
It is a pleasure to thank Halina Abramowicz and Lonya Frankfurt for 
giving me some better insight of the interplay between soft and hard 
interactions. Special thanks to Giullio d'Agostini and his committee for 
organizing such an excellent workshop.


\end{document}